# Controlling a Software-Defined Network via Distributed Controllers

Volkan Yazıcı[1], M. Oğuz Sunay[1], Ali Ö. Ercan[1]

[1]Özyeğin University, Istanbul, Turkey

E-mail: [1]{volkan.yazıcı, oguz.sunay, ali.ercan}@ozyegin.edu.tr

*Abstract:* **In this paper, we propose a distributed OpenFlow controller and an associated coordination framework that achieves scalability and reliability even under heavy data center loads. The proposed framework, which is designed to work with all existing OpenFlow controllers with minimal or no required changes, provides support for dynamic addition and removal of controllers to the cluster without any interruption to the network operation. We demonstrate performance results of the proposed framework implemented over an experimental testbed that uses controllers running Beacon.**

## 1 INTRODUCTION

Today's networks have become exceedingly complex, because they implement an ever increasing number of distributed protocols standardized by IETF and the individual packet routing/switching components within these networks use closed and proprietary programs that take these protocols into account. In this environment it is too difficult, if not impossible, for network operators, third parties, including researchers, and even vendors to innovate [1]. To address this problem, in 2011, the Open Networking Foundation (ONF) has been formed with the aim of promoting a new networking paradigm, called *Software-Defined Networking* (SDN). The fundamental idea behind SDN is a network architecture where the control plane is decoupled from the data plane. This abstraction opens up the possibility for a programmable network, where the administrators can customize the network to fit their needs.

The *OpenFlow* communication protocol is one of the enablers of the SDN paradigm [1]. It provides a common set of instructions for the control plane to interact with the data plane realized via packet-forwarding hardware. The control-plane, which is commonly referred to as the controller or the network operating system in SDN, resides on a dedicated server and commands the packet-forwarding hardware through the OpenFlow protocol as illustrated in Figure 1. This abstraction enables the controller to easily enforce flow-based sophisticated traffic management policies (routing, QoS, VLAN tagging, etc.) in the network.

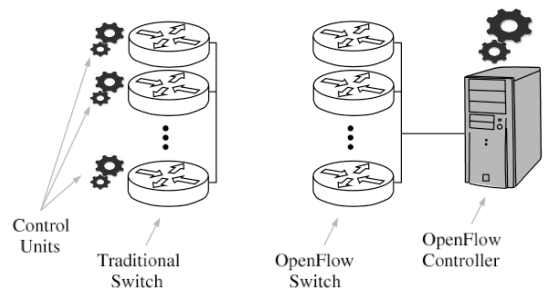

**Figure 1: Comparison of traditional and SDN networks**

Recent studies conducted on the networks of many real-world data centers showed that such networks necessitate the handling of about 150 million flows per second [2]. On the other hand, today's OpenFlow controllers are known handle at most 6 million flows per second on a high end dedicated server with 4 cores (see Section III.) Therefore, implementation of SDN for one of such data center networks requires a controller running either on an appropriate mainframe computer with sufficiently many cores or a server cluster where each server is composed of limited cores.

Implementation of the controller on a cluster offers a number of benefits. First, this platform is scalable, as an increasing load on the controller is easily handled by introducing new servers to the cluster. Second, the cluster offers more reliability than an implementation on a single mainframe, which presents a single point of failure. For this reason, we propose a cluster based distributed OpenFlow controller framework as illustrated in Figure 2 in this paper.

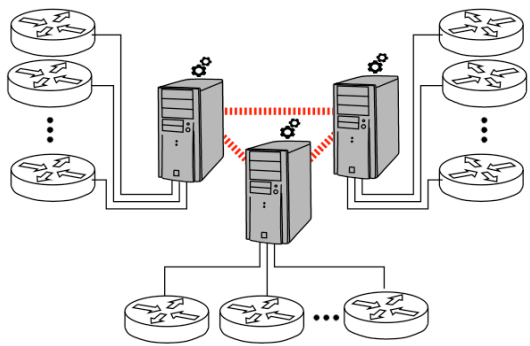

**Figure 2: Distributed OpenFlow controller architecture**

This work has partially been supported by European Commission Grant number PIRG06-GA-2009-256326 and Turk Telekom Grant number 11580-04.

**Corresponding author:** Volkan Yazıcı, Özyeğin University, Çekmeköy, İstanbul, Turkey, +90(216)564-9000, volkan.yazici@ozu.edu.tr

It is well-known that coordination is an important pillar of distributed systems. For the proposed cluster based distributed controller architecture, a well-designed coordination framework is necessary to allow for load balancing between the distributed controllers and for replacement of failed controllers by active ones so that scalability and reliability is sustained with zero network down-time.

A number of papers have recently appeared in the literature on distributed implementations of OpenFlow controllers. In [3], the authors present a distributed NOX-based controllers interwork through extended GMPLS protocols. [4] proposes to deploy multiple instances of the same NOX controller on a set of distributed nodes. [5] proposes to use a distributed set of autonomous controllers, but does not provide means for switch migration amongst them. In [6], a new platform is introduced, over which a distributed network control plane may be implemented.

In this paper, we propose a distributed controller framework and an associated coordination framework that achieves scalability and reliability even under heavy data center loads. The proposed architecture provides two novel key features that are not present in previous related work. First, the proposed framework provides support for dynamic addition and removal of controllers to the cluster without any interruption to the network operation. Second, the proposed distributed controller and associated coordination framework is designed to work with all existing OpenFlow controllers with minimal or no required changes.

The remainder of this paper is organized as follows. In Section II, we introduce the coordination framework for the cluster based controller implementation. We present experimental results of the proposed framework in Section III and conclude in Section IV.

## 2 PROPOSED FRAMEWORK

In this paper, we consider an OpenFlow controller implementation that is cluster based as illustrated in Figure 2. In the implementation, multiple controllers are realized in a cluster, each on a distinct server. The proposed framework may use any of the operating systems from the literature to implement the controllers, with simple modifications, if necessary.

In the proposed framework, controllers in the cluster communicate with each other using the JGroups membership notifications and messaging infrastructure [7]. JGroups is a mature, robust and flexible group communication library used in many data centers for various mission critical applications. In this setting, the controllers elect a master node amongst them which conducts and maintains the global controller-switch mapping in the network. The master node is periodically monitored by all other nodes, and if it is found to be inaccessible, it is immediately replaced by one of the other nodes. Thus, the proposed framework does not expose a single-point-of-failure.

The proposed controller architecture interfaces the switches in the network as well as the applications running above posing as a single, centralized controller. In other words, the switches and the applications are unaware of the switch assignments and re-assignments in the network to the individual controllers in the cluster. This provides a seamless, compatible operation with legacy OpenFlow switches and applications. However, if a switch and/or application is aware of the underlying distributed architecture, the necessary API to utilize its features is also provided.

We now discuss how the master controller is selected, how the switches in the network are mapped to the controllers, how load balancing between the controllers is achieved, and what happens when one of the controllers becomes inaccessible in the proposed framework. Next, we present a network model to simplify the discussion.

### 2.1 Notation

Let $C=\{c_1,\ldots,c_n\}$ denote the controllers in the cluster and $S=\{s_1,\ldots,s_m\}$ denote the switches in the network. The network is represented by an undirected graph $G=(V,E)$, where $V$ denotes the vertices and $E$ denotes the edges. Vertices are composed of controllers and switches, i.e, $V=C \cup S$ and edges are composed of two-tuples representing the connections between vertices, that is, $E \subseteq \{(v_k,v_l) | v_k,v_l \in V\}$, $k,l=1,2,\ldots,n+m$. $k$ and $l$ are omitted in the text from now on for brevity. The switch-controller mapping is given by the set $M$, which is a subset of $E$, connecting vertices between $C$ and $S$; $M \subseteq \{(c_k,s_l) \in E \mid c_k \in C, s_l \in S\}$. In $M$, each switch is constrained to be connected to a single controller. That is, if $s_l \in S$ and $c_p,c_q \in C$, then $(c_p,s_l) \in M$ and $(c_q,s_l) \in M$ if and only if p=q.

In an OpenFlow network, controllers and switches are connected via an IP network. In a given network $G=(V,E)$, an IP network is represented by a graph $N=(V_N, E_N)$, where vertices and edges are given by $V_N \in V$ and $E_N=\{(v_k,v_l)|v_k,v_l \in V_N\}$, respectively. Here, each edge $(v_k,v_l) \in E_N$ is constrained to have a path from $v_k$ to $v_l$ in $G$. That is, $(v_k,v_l) \in E_N$ for $v_k,v_l \in V_N$ if and only if $\{(v_k,v_{i1}),(v_{i2}, v_{i3}),\ldots,(v_{ir},v_l)\} \subseteq E$ and $\exists$ $v_{i1},v_{i2},\ldots,v_{ir} \in V$. In an IP network $N$, each vertex $v_i \in V_N$ is assigned a set of IP addresses denoted by $IP_N\{v_i\}$ (assigning multiple IP addresses to a single interface is possible through IP aliasing). Assigned IP addresses in $N$ are chosen to be pairwise disjoint, that is, $IP_N\{v_k\} \cap IP_N\{v_l\} = \emptyset$ for $k \neq l$ and $\forall$ $v_k,v_l \in V_N$.

### 2.2 Local IP Networks

In the proposed framework, the network of controllers and switches is divided into two distinct IP networks: an IP network $A=(V_A,E_A)$ for the controller-controller communication (i.e., $V_A=C$) and an IP network $\mathcal{B}=(V_B, E_B)$ for the controller-switch communication (i.e., $V_B=C \cup S$).

IP network $\mathcal{A}$: Here, each controller $c_i \in V_A$ is statically assigned a unique IP address $C_i$, i.e., $IP_A\{c_i\}=\{C_i\}$.

IP network $\mathcal{B}$: Here, a unique IP address $S_i$ is assigned to each switch $s_i \in S$, i.e., $IP_B\{s_i\}=\{S_i\}$. In this network, a pool of IP addresses, $P_i$ describes the controllers. There are as many $P_i$'s as there are switches in this pool. At a given time, each switch, $s_i$ is statically configured to connect to the controller with IP address $P_i$. The master controller decides on how the switches are mapped to the controllers by partitioning the IP address pool $P_i$ amongst all controllers, including itself. Once a switch $s_i$ is mapped to controller $c_m$, the IP address $P_i$ is dynamically assigned to that controller by IP aliasing. Thus at any given time, if $P_l \in IP_B\{c_k\}$, controller $c_k \in C$ is said to be controlling the switch $s_l \in S$. If, for some reason, the controller-switch mapping changes at some stage, the IP alias $P_i$ is moved to the new controller that starts to control the switch. To avoid multiple controllers trying to control the same switch or a switch not being controlled by any of the controllers, IP addresses are assigned to be mutually exhaustive and pairwise disjoint, i.e., $\{P_i\}=\cup_{ck} IP_B\{c_k\}$ and $IP_B\{c_k\} \cap IP_B\{c_l\} = \varnothing$ for $c_k, c_l \in C$ and $k \neq l$.

Figure 3a shows a sample OpenFlow network composed of 2 controllers ($c_1$, $c_2$) and 5 switches ($s_1,\ldots,s_5$). Here, the initial mapping between the controllers and the switches is given by $\mathcal{M}=\{(c_1,s_1), (c_1,s_2), (c_2,s_3), (c_2,s_4), (c_2,s_5)\}$. Switches $s_1$ and $s_2$ are controlled by controller $c_1$, i.e., $IP_B\{c_1\} = \{P_1,P_2\}$, $IP_B\{s_1\}=\{S_1\}$ and $IP_B\{s_2\}=\{S_2\}$. Similarly, switches $s_3,s_4,s_5$\$ are controlled by controller $c_2$ i.e., $IP_B\{c_2\} = \{P_3,P_4,P_5\}$, $IP_B\{s3\}=\{S_3\}$, $IP_B\{s_4\}=\{S_4\}$ and $IP_B\{s_5\}=\{S_5\}$.

### 2.3 Master Controller Selection

In the proposed framework, each controller in the cluster is equipped with the same algorithm that generates and updates the network mapping. In this setting, a master node is responsible for realizing the controller-switch mapping updates. The master controller is determined using a distributed atomic integer primitive provided by JGroups. This scheme is outlined in Algorithm 1.

---
**Algorithm 1** REPLACEMASTER()
1: **repeat**
2:    $c_{prev} \leftarrow$ MASTER.GET()
3:    $c_{next} \leftarrow$ FINDMASTER()
4: **until** MASTER.COMPAREANDSWAP($c_{prev}, c_{next}$)

---

There are various approaches to the selection of a master in a cluster environment [8]. These studies generally employ a cost function with a set of user provided constraints over the measurements collected throughout the system. Then, the most effective configuration is selected from all available candidates. In the proposed framework, the processing cost imposed by the cluster is almost negligible on a master controller and is rarely encountered, e.g., while balancing loads or in case of a server failure. In the framework presented herein, when FINDMASTER() is invoked, it is set to return the controller with the smallest system load. To avoid frequent master changes, this algorithm is invoked only when the current master becomes unsuitable based on some user-defined criteria.

In the proposed framework, initially, the controller that first completes the execution of algorithm 1 is established as the master. In this algorithm, when a controller decides that the master node needs to be changed, it finds a suitable node for replacement (line 2 and 3). Then, using the atomic integer primitive COMPAREANDSWAP(), this algorithm repeatedly tries to replace the master node, until it succeeds to do so (line 4) or number of repetitions exceed a certain threshold. Once the master controller is replaced, JGroups ensures that it gets atomically propagated throughout the cluster. Each cluster member checks if it is the master node, if so, it starts executing the regular switch-controller mapping checks and decisions.

Therefore, to avoid making master node a single point of failure, the rest of the controllers in the cluster regularly check the working status of the master node and, in case of a failure, attempt to replace it.

### 2.4 Mapping

In the proposed framework, the master controller dynamically partitions the IP addresses Pi amongst all controllers. A controller $c_k$ controls a switch $s_l$ by adding IP address $P_l$ into $IP_B\{c_k\}$, i.e., its list of IP aliases in network $\mathcal{B}$. This necessitates generation and updating of the mapping $\mathcal{M}$ regularly. The mapping information is stored locally by all controllers and this database is automatically synchronized amongst all of them via JGroups with every local update.

The mapping $\mathcal{M}$ is generated and updated by the master controller using the GENERATEMAPPING() function. This function takes the two IP networks $\mathcal{A}$, $\mathcal{B}$, current mapping $\mathcal{M}$ (empty set if no current mapping exists) and a set of system statistics parameters, STATS as input parameters. STATS is composed of statistics such as link traffic, controller loads, etc., which are collected and provided by the controller architecture. Additionally, a network administrator might decide to route certain flows over certain machines for security concerns, custom applications might require customized quality-of-service measures, etc. All such constraints need to be taken into account in the GENERATEMAPPING() function. The implementation of the GENERATEMAPPING() function highly depends on the work flow of the underlying network and is out of the scope of this work.

The result of the GENERATEMAPPING function is propagated throughout the controller cluster by the master controller. For this purpose, the master controller invokes the SETMAPPING($\mathcal{A},\mathcal{B},\mathcal{M}_{old},\mathcal{M}_{new}$) algorithm, where $\mathcal{M}_{old}$ and $\mathcal{M}_{new}$ arguments denote the the old and the new mappings, respectively. In Algorithm 2, SETMAPPING() function is detailed. Here, for every switch the function determines its current (line 2) and next (line 3) controller, implied by the mappings $\mathcal{M}_{old}$ and $\mathcal{M}_{new}$, respectively. Next, the function decides on how to realize the operation of migrating a switch from $c_{old}$ to $c_{new}$ using COALESCE() (line 4). Here, if $c_{old}$ is alive, it is selected as the first point of contact for the mapping update message. If not, $c_{new}$ is selected as the first point of contact. Finally, in order to trigger the actual switch migration, the function makes a remote procedure call on the first point of contact, $c_r$ to run the function MOVE() (line 5).

---
**Algorithm 2** SETMAPPING($\mathcal{A}, \mathcal{B}, \mathcal{M}_{old}, \mathcal{M}_{new}$)
1: **for** $\ell \leftarrow 1 \ldots m$ **do**
2:    $c_{old} \leftarrow \exists c_k$ for $(s_\ell, c_k) \in \mathcal{M}_{old}$
3:    $c_{new} \leftarrow \exists c_k$ for $(s_\ell, c_k) \in \mathcal{M}_{new}$
4:    $r \leftarrow$ COALESCE($c_{old}, c_{new}$)
5:    $c_r$ ! MOVE($\mathcal{A}, \mathcal{B}, s_\ell, c_{old}, c_{new}$)
6: **end for**

---

Whenever a controller receives a remote call to run the MOVE() function, it is expected to either release a switch for some other controller to subsequently acquire it, or acquire an already released switch. This operation is detailed in Algorithm 3. Here, the function first inquires its own rank in the cluster (controller ID) (line 1) and determines if it is expected to release (line 2) or acquire (line 6) the switch. If it is invoked to release the switch, first it releases the control of the switch (line 3) and updates its list of IP addresses in network $\mathcal{B}$ (line 4). Then, it invokes the controller that will take control of the switch ($c_{new}$) (line 5) to run the MOVE() function. Otherwise, if MOVE() is invoked at the controller to acquire the switch, it first acquires the control of the switch (line 7) and updates its list of IP addresses in network B (line 8). Finally, the acquiring controller alerts the switch to reset its ARP cache (line 9).

---
**Algorithm 3** MOVE($\mathcal{A}, \mathcal{B}, s_\ell, c_{old}, c_{new}$)
1: $r \leftarrow$ CURRENTRANK()
2: **if** $c_r = c_{old}$ **then**
3:    RELEASE($P_\ell$)
4:    IP$_\mathcal{B}(c_r) \leftarrow$ IP$_\mathcal{B}(c_r) \setminus P_\ell$
5:    $c_{new}$ ! MOVE($s_\ell, c_{old}, c_{new}$)
6: **else if** $c_r = c_{new}$ **then**
7:    ACQUIRE($P_\ell$)
8:    IP$_\mathcal{B}(c_r) \leftarrow$ IP$_\mathcal{B}(c_r) \cup P_\ell$
9:    ARPPING($s_\ell$)
10: **end if**

---

## 2.5 Operation

The master controller regularly observes the network statistics provided by STATS. If a load imbalance (induced by controller/switch addition, controller/switch failure, flow/traffic changes, \etc.) is detected, the master first triggers a GENERATEMAPPING() call. The resulting mapping is then updated for each controller via JGroups and is executed by the SETMAPPING() function. The operation of the proposed framework is examplified in Figures 3b and 3c. Here, assume that the controller $c_2$ is overloaded. In the new mapping, the switch s3 is reassigned to $c_1$ by GENERATEMAPPING() to alleviate the load of $c_2$. The IP address $P_3$ is first released by $c_2$ and subsequently acquired by c1 through executions of SETMAPPING() first on $c_2$ and then on $c_3$. In the new configuration, $c_3$ is no longer overloaded and the loads of $c_1$ and $c_2$ are more even. The transition is seamless for the switch $s_3$ since it is still connected to IP address $P_3$ for controller traffic.

When a new controller is added to the cluster, the rest of the cluster is instantly notified by the JGroups membership notification feature. Consequently, a new GETMAPPING()-SETMAPPING() cycle takes place so that some of the loads of the existing controllers is passed on to the newly added one to provide some desired level of load balancing. Similarly, if a controller fails or is selectively turned off by the system administrator due to low traffic, the rest of the cluster is immediately notified by JGroups and again, a new GETMAPPING()-SETMAPPING() cycle is executed. Newly computed mapping will replace the inaccessible controllers with the working ones. This is exemplified in Figure 3d. Here, $c_2$ dies. Therefore in the new mapping all switches are assigned to $c_1$.

## 2.6 Routing

The use of JGroups facilitates synchronism of the network map $\mathcal{M}$ across all controllers. Therefore, even though the switches in the network are distributed dynamically across the controllers, the framework allows for optimized end-to-end routing of flows over the entire network. The necessary messaging to facilitate the routing operation is beyond the scope of this paper.

## 3 RESULTS

We have conducted experiments to assess the performance of the distributed controller framework proposed herein. In the experiments, we implemented the framework using Beacon [9] Beacon has a successful track record in performance benchmarks (see Figure 4a) and enables the use of JVM libraries.

To run the experiment, we first enhanced Beacon by adding a new OSGi bundle, called `cluster`. Four controller machines - running Debian GNU/Linux 6.0.4 (i686) on a system with Dual-Core 2.80GHz CPU, 2GB

RAM, and JDK 1.6.0-30 - were configured with the enhanced Beacon. All machines were connected through an unmanaged gigabit switch and cluster communications were forced to run on physically separate NICs on each machine.

We first examine the cluster power up using the experimental setup. We observe that it takes approximately 12 sec. for the cluster OSGi bundle to start up initially when no other controllers are active. If there is at least one active controller in the cluster, the start up time is approximately 3 sec. The 9 sec. difference is due to discovery time during JGroups channel initialization. In the proposed framework, we measure that it takes in the order of under 50 milliseconds for members to get notified by the removal/arrival of a member in the cluster.

In the proposed framework, controllers acquire and release switches via maintenance of IP aliases. Using the experimental setup, we measure the time it takes for one of the controllers in the cluster to acquire and release IP aliases. The results are plotted in Figure 4b. We observe that the time it takes for these operations is slightly convex. However, we note that even with 254 simultaneous IP alias changes, the operation clocks under 5 sec.

Next, we measure the time it takes to migrate a group of switches from one controller to another. The results are plotted in Figure 4c. We note here that the switch migration includes an IP alias location change as well necessary communications over JGroups and kernel calls. In the figure, we observe that even 254 simultaneous switch migrations clock around 8 sec. approximately 4 of which is due to IP alias relocate operations. We note here that under normal network operations, the number of simultaneous switch migrations would be significantly less than 254.

Finally, the experimental setup with 4 controllers and 4 emulated switches is used to assess the performance increase with multiple controllers. Switch emulators are configured to run cbench [10] instances to stress the controller throughput for a period of ten seconds. Each stress test is repeated ten times. The first and the last runs are discarded to remove the effects of warm-up and cool-down times. The average number of controller responses per second per switch when one, two, three or four controllers are used are reported in Figure 4d. When multiple controllers are used, the switches are assigned to the controllers in a load balanced manner. As seen in the figure, the number of responses per second per switch increase super-linearly as more controllers are used. This is because, the possible combinations of interaction between the switches assigned to a controller increases quadratically with the number of switches. Thus, the overhead of coordinating switches per controller increases super-linearly with the number of switches assigned to it. This results in the super-linear increase in the performance as less switches are assigned to a controller.

## 4 CONCLUSIONS

This paper presents a distributed OpenFlow controller architecture and an associated coordination framework that achieves scalability and reliability even under heavy data center loads via the use of JGroups. The proposed architecture, which is designed to work with all existing OpenFlow controllers with minimal or no required changes, provides support for dynamic addition and removal of controllers to the cluster without any interruption to the network operation. Experimental results confirm that using the proposed framework, migration of switches amongst multiple controllers, addition/removal of controllers to the network is possible. The use of JGroups facilitates synchronism of the network map across all active controllers. As such, this framework allows for optimized end-to-end routing of flows across the network while achieving scalability and reliability of the network controllers.

## References

[1] N. McKeown et al., "OpenFlow: Enabling Innovation in Campus Networks," *SIGCOMM Comput. Commun. Rev.* vol. 38, pp. 69-74, March 2008.
[2] T. Benson, A. Akella and D.A. Maltz, "Network Traffic Characteristics of Data Centers in the Wild," in *Proc. ACM IMC*, New York, NY, USA, 2010.
[3] R. Martinez et al., "OpenFlow-Based Hybrid Control Plane witin the CTTC ADRENALINE Testbed," *in Proc. OFELIA Workshop*, Geneva, Switzerland, September 2011.
[4] A. Tootoonchian and Y. Ganjali, "HyperFlow: A Distributed Control Plane for OpenFlow Networks," *in Proc. INM/WREN*, San Jose, CA, USA, April 2010.
[5] C. Macapuna, C. Rothenberg, and M. Magalhaes, "In-Packet Bloom Filter Based Data Center Networking with Distributed OpenFlow Controllers," *in Proc. IEEE Globecom*, December 2010.
[6] T. Koponen et al., "Onix: A Distributed Control Platform for Large-Scale Production Networks," *in Proc. USENIX OSDI*, Vancouver, BC, Canada, October 2010.
*[7]* B. Ban, "Design and Implementation of a Reliable Group Communication Toolkit for Kava," *http://www.jgroups.org/papers/Coots.ps.gz*.
[8] G. Shau, F. Berman, and R. Wolski, "Master/Slave Computing on the Grid," *in Proc. IEEE 9th Heterogeneous Computing Workshop*, Washington DC, USA, 2000.
[9] D. Erickson, "Beacon: A Fast, Cross-Platform, Modular, Java-Based OpenFlow Controller," *http://beaconcontroller.net/,* 2011.
[10] C. Rostos et al. "OFLOPS: An Open Framework for OpenFlow Switch Evaluation," *in Proc. PAM*, 2012.

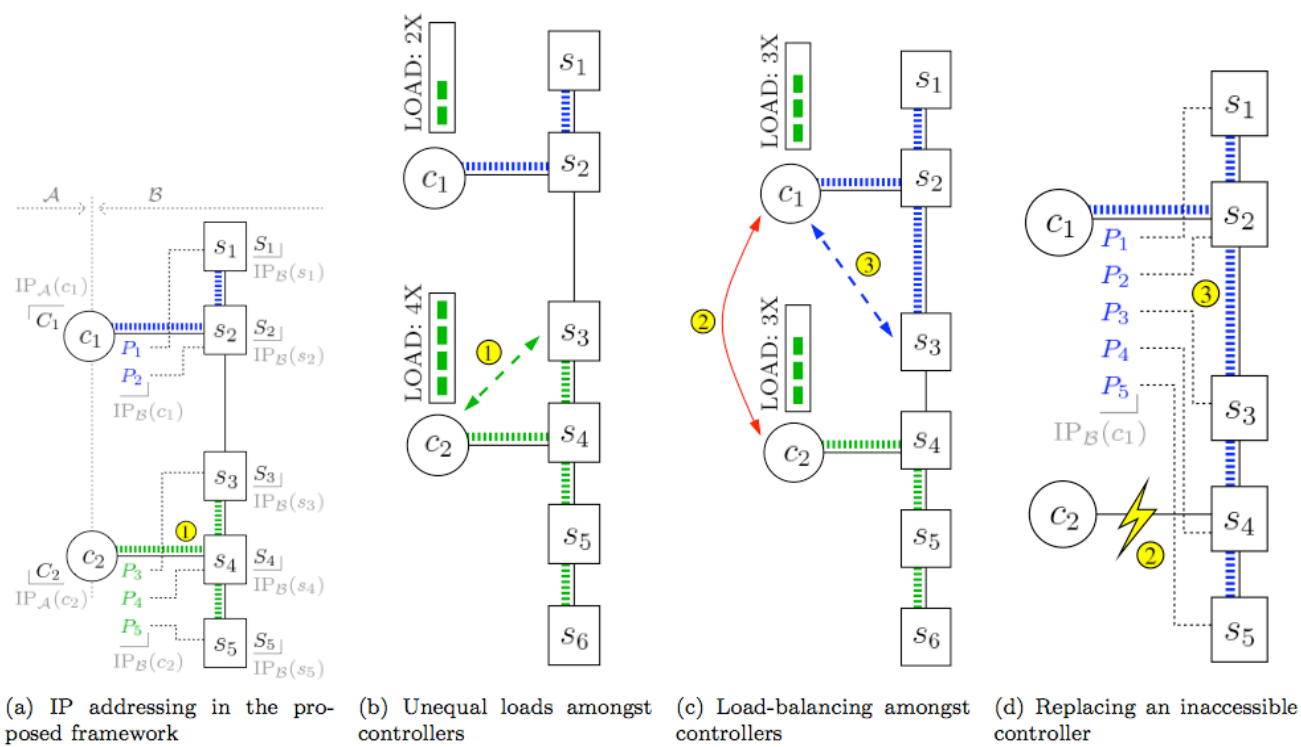

Figure 3: Controller coordination using the proposed framework

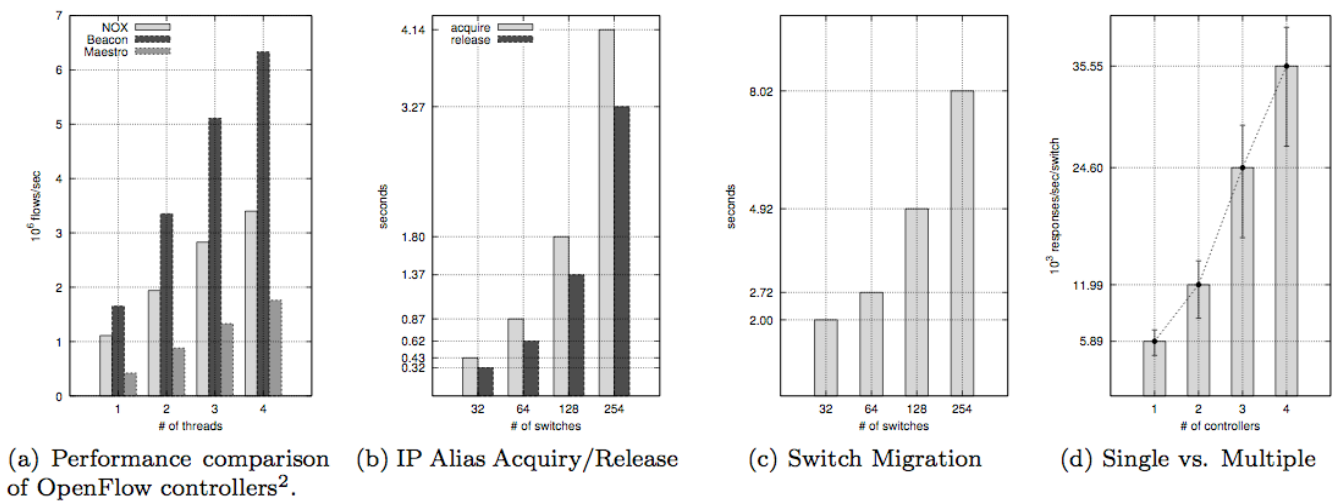

Figure 4: Experimental results for the proposed distributed OpenFlow Controller Framework